\documentclass[twocolumn,showpacs,amsfonts,amssymb,pre,floatfix]{revtex4-1}
\usepackage{graphicx,epsfig}
\usepackage{amsmath}
\usepackage{hyperref}
\usepackage{color}

\begin{document}
\title{International Transmission of Shocks and Fragility of a Bank Network}
\author{Xiaobing Feng}
\affiliation{School of Finance, Shanghai University of International
Business and Economics, Shanghai 201612, P.R. China}
\affiliation{Center of Financial Engineering, Shanghai Jiaotong University, Shanghai 200022, P.R. China}
\author{Woo Seong Jo}
\affiliation{Department of Physics and BK21 Physics Research Division,
Sungkyunkwan University, Suwon 440-746, Korea}
\author{Beom Jun Kim}
\email[Corresponding author: ]{beomjun@skku.edu}
\affiliation{Department of Physics and BK21 Physics Research Division,
Sungkyunkwan University, Suwon 440-746, Korea}
\begin{abstract}

The weighted and directed network of countries based on the number of overseas
banks is analyzed in terms of its fragility to the banking crisis of one
country. We use two different models to describe transmission of shocks, one
local and the other global.  Depending on the original source of the crisis,
the overall size of crisis impacts is found to differ country by country. For
the two-step local spreading model, it is revealed that the scale of the first
impact is determined by the out-strength, the total number of overseas branches
of the country at the origin of the crisis, while the second impact becomes
more serious if the in-strength at the origin is increased.  For the global
spreading model, some countries named ``triggers''  are found to play important
roles in shock transmission, and the importance of the feed-forward-loop mechanism is pointed
out. We also discuss practical policy implications of the present work.

\end{abstract}


\maketitle
\section{Background}

  The increasing globalization provides important advantages in terms of
risk sharing and risk diversification in banking and financial markets, but
it also facilitates the risk spreading among different nations~\cite{helbing}. The recent
financial crisis or shock, originated from United States (US) and spread to other countries,
has witnessed that many overseas banks cut back their loans to the local
markets and withdrew their representations. The withdrawals cause the
liquidity shortage of the host countries which then lead to subsequent
withdrawals of the host countries' overseas branches in the crisis attacked
countries. This process can continue and the shocks in this situation are
hence defined as sequential shocks in this study.

There are abundant literature on transmission mechanism of shocks in global
banking: \cite{peek} examines how Japanese asset bubbles have been transmitted
to US via lending responses of Japanese overseas banks in US in the
80's;
\cite{schnabl} explores the mechanism of 1998 Russian debt default as a
negative liquidity shock to international banking and its impacts on the
banking of Peru; \cite{popov} studies the effect of financial distress in
foreign parent banks on the local SME (small and medium enterprises) financing
in 14 central and eastern European countries during the early stage of the 2007
financial crisis.

Within the framework of complex network theory, this paper introduces
models for shock transmission and examines the different impacts of the sequential
shocks on the global banking network. We
find that the damage of the shocks is closely related to the directionality of
the edges and the network topology plays an important role.

\begin{figure*}
\includegraphics[width=0.98\textwidth]{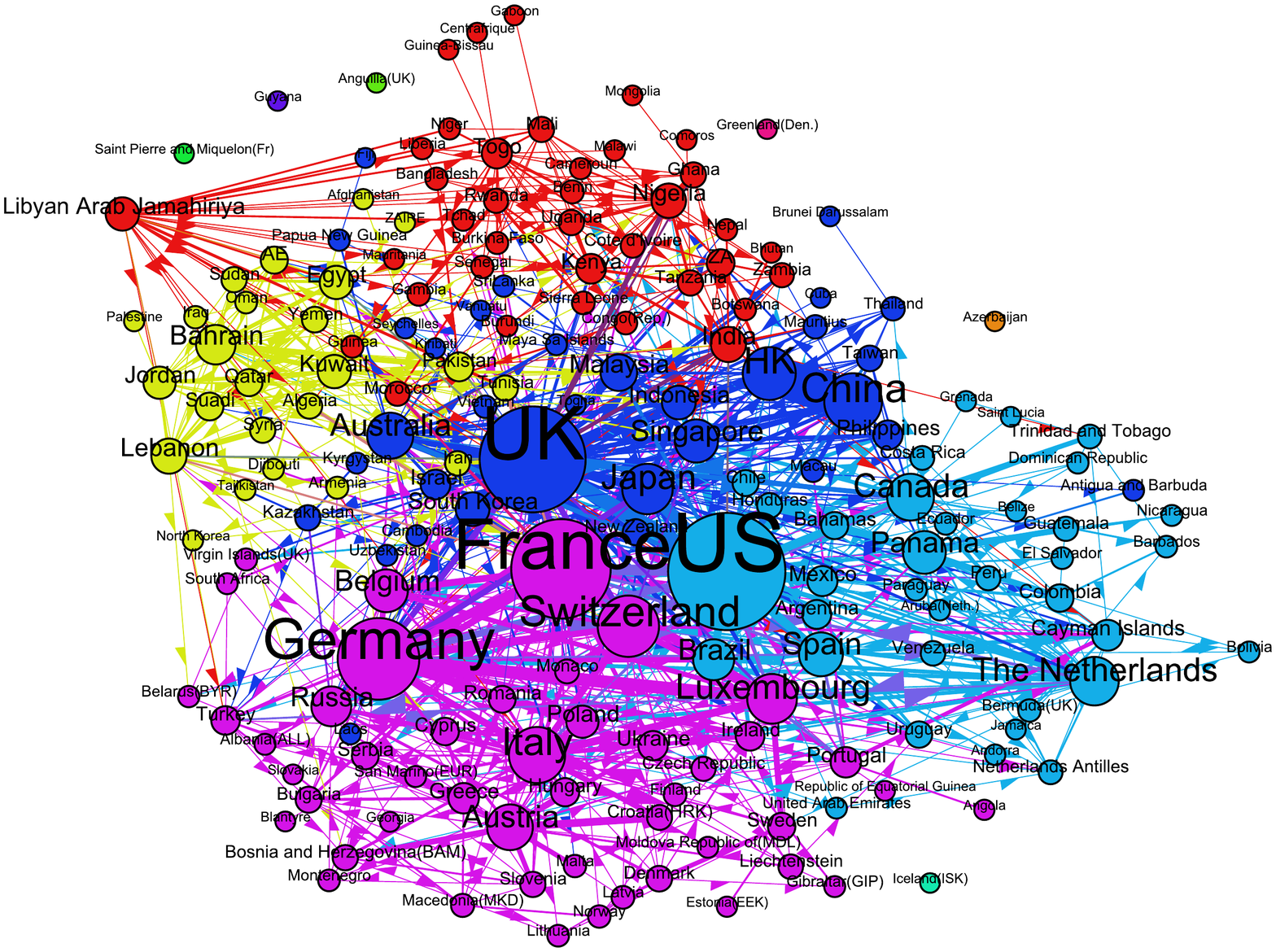}
\caption{(Color online) The international bank network drawn by using
the software Gephi. The sizes of vertices are proportional to
the sum of the out-strength and the in-strength.
Community detection algorithm~\cite{blondel} in Gephi yields
clusters and we represent them by different colors.
Note that geographic locations of countries and community structures are closely related to each other.}
\label{fig:network}
\end{figure*}

\section{Literature Review on Complex Network Methodology}
The research using complex system approach in banking has also experienced
three stages of the development. In the earlier stage, \cite{allen} claims that
the possibility of contagion effect depends on the structure of interbank
linkages. They believe that a ``complete structure of linkages'' will share the
risk more easily than ``an incomplete structure'', hence the risk sharing
effect. \cite{freixas,*upper} further considers a structure of uni- and
multi-money center banking systems, where the banks on the periphery are linked
to the bank at the center but not to each other. Multi-tiered banking system
has then been examined with the similar approach. These researches have shown that
scholars have started to notice the banking structure has impacts on contagion
even though the models are simple and has not yet formally introduced the
complex system theory. The second stage is symbolized by a conference entitled
``New Directions for Understanding Systemic Risk'', which brought together experts
from various disciplines to explore parallels between systemic risk in finance
and systemic risk in engineering, ecology, and other fields of
sciences~\cite{kambhu}. After that a series of interdisciplinary researches
using various modeling and theory have emerged~\cite{soramaki,*beyeler,*iori}.
The 2007 US financial crisis becomes another driving force of this line of the
research. At this stage researchers have started to apply complex system theory
to disclose topology and features of different financial markets such as
international trade network, investment network, interbank clearing
network~\cite{may,*bech,*gai,*towe,*feng,*garas}.

There are three issues in the current literature that need to be addressed and can be improved: first, the current research has focused on the disclosure of topology of the financial network, the study on dynamics and the interaction between the dynamics and topology are limited; second, the application of knowledge from both economics and complex system is rather mechanic; third, the empirical research has mainly emphasized on two markets, interbank and payment systems, where data are relatively easy to obtain. Other markets, such as a global banking network,
have been hardly researched, hence the focus of this research.

In this paper, ``fragility'' is defined as the reduction of the number of banks
as a result of exogenous shocks of different sequence and different sizes. The
initial shock will generate rounds of sequential shocks at later stages to
propagate throughout the entire network. The initial shock starts from the
shock targeted country, such as US in 2007; it then spreads to the hosting
countries where US has its overseas banks; resulting in more reduction of
banks. A financial crisis in this paper hence is a breakdown of the network’s
bank linkages, a collapse of all or part of the network structure.

\section{Global Bank Network}
\subsection{The Database }
   The data used in this research is from Bankscope which has information on
over 30,000 public and private banks throughout the world from 2009 to 2011.
Each bank report contains detailed consolidated and unconsolidated balance
sheet and income statement. Data comes from Fitch Ratings and six other
sources. It also provides company and country risk ratings and reports,
ownership, and security and price information. This database is produced by
Bureau van Dijk.

	 The information in the database that can be used for network construction
are the number of branches and subsidiaries that each parent bank has
established overseas. Subsidiaries are banks that are completely or partly
owned and wholly controlled by parent banks that owns more than half of the
subsidiary's stock. For the purpose of this research subsidiary banks and branch
banks are equally treated as overseas banks. The information on the overseas
location of the two types of banks can be found in ``ownership'' category of
the database.

\subsection{The Topology of the Global Banking Network}

By using gathered information, we construct the directed and weighted network of $N=182$
countries, in which the directed arc from the country $i$ to country $j$ is assigned
the weight $w_{ij}$ that is the number of banks $i$ puts in $j$. We in the present work
focus on international banks and thus domestic branches of banks are disregarded, i.e.,
$w_{ii} = 0$. The number of arcs
that has nonzero weights amounts to $M = 1055$ (and thus the average out- and in-degree 5.80),
which indicates that the network is very sparse since the total number of possible arcs
is $N(N-1) = 32942$, much larger than $M$.

The network in Fig.~\ref{fig:network} is drawn by using the software Gephi.
Although sparse, the network is well connected: the giant component size
is comparable to the system size and there are 6 isolated very small islands.
The top five
countries in terms of vertex strengths in global network are United Kingdom (UK), US, France,
Germany  and Switzerland in this order. The top five countries in terms of
vertex strengths in Asia are Japan, China, Hong Kong (HK), Singapore, and South
Korea in this order. In terms of in-degree the top five countries are UK, US,
Russia, Switzerland, France. It is interesting to note that Panama island is
ranked the six as one of the famous offshore banks. The top five countries in
Asia in terms of in-degree, are Singapore, HK, China, Japan and South Korea.
In terms of
out-degree the top five countries are UK, US, France, Germany and the
Netherlands, and in Asia they are Japan, China, South Korea, HK and
Singapore. As historical Asian international financial center, Singapore, Japan
and HK are still playing important roles, and China is gaining its
momentum.

  In general, the network shows that in-degree of a country is different
from its out-degree. It implies that the export of banking services of a
nation is not equal to the imports of banking service of the country. Hence
service trade imbalances are observed in the global banking network.
We identify ten clusters by using Gephi which uses the community detection
method in~\cite{blondel} and display them in different colors in Fig.~\ref{fig:network}.
It needs to be mentioned that even in this era of global finances, we still observe the
strong dependence on geographic locations of countries. In Fig.~\ref{fig:network}, the
different communities have different colors and it is clearly seen that each community
is closely linked to each continent.
Interestingly, our finding is in contrast
to the argument in the literature that geographic distances do not play
important roles with the development of electronic banking~\cite{krizek}.

\begin{figure*}
\includegraphics[width=0.98\textwidth]{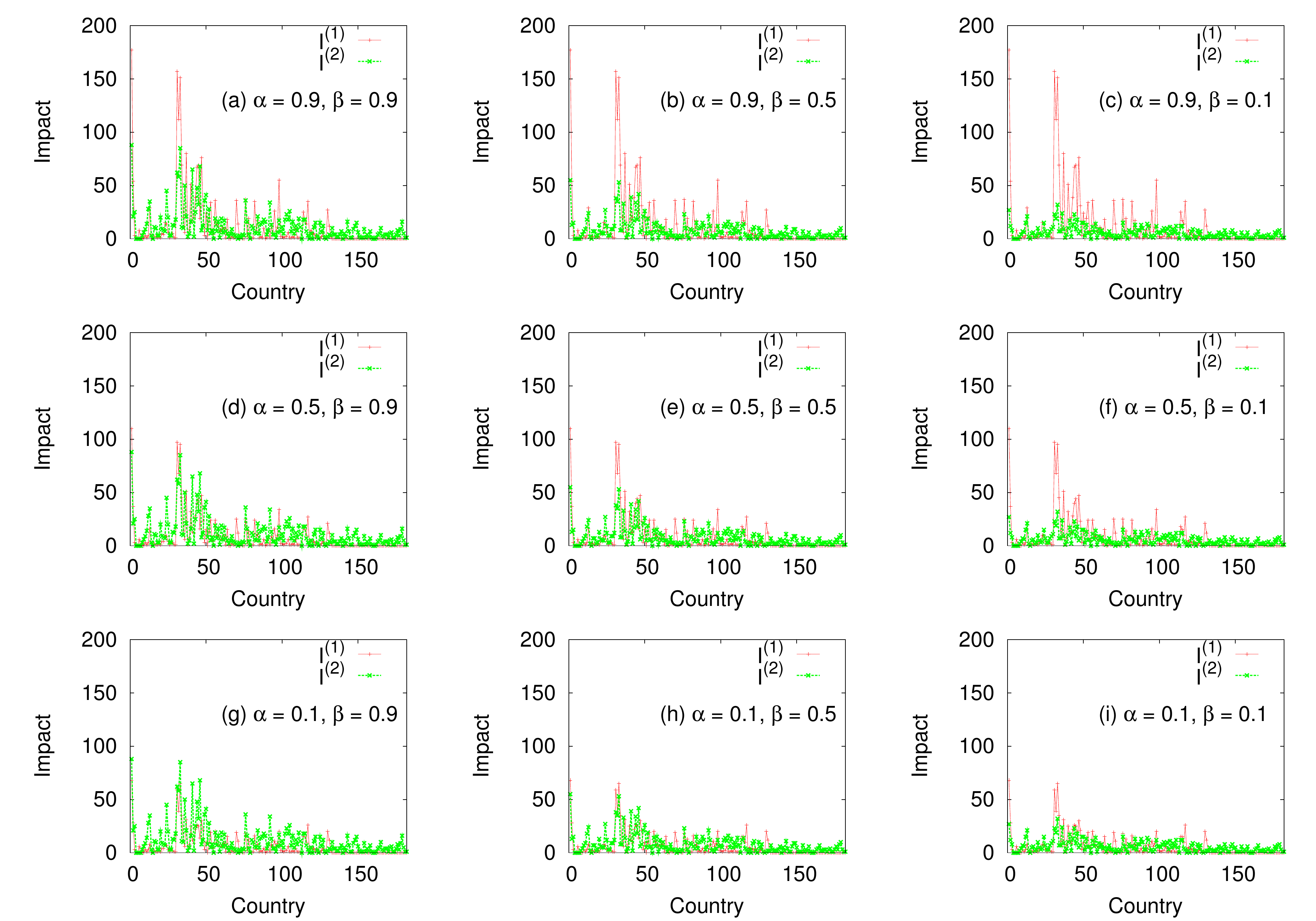}
\caption{(Color online) Local spreading model: The first impact and the second impact measured by the
change of total number of foreign banks is shown for different values of $\alpha$ and $\beta$.
As $\alpha$ and $\beta$ become larger (i.e., the seriousness of the first and the
second waves become stronger), the impact sizes also increase. The top 10 countries
which cause the most serious crisis in the first and the second waves are listed
in Table~\ref{tab:iks}
}
\label{fig:local}
\end{figure*}


\begin{table*}
\caption{Local spreading model: The list of the top 10 countries that caused the most serious
first and the second impacts for $\alpha = \beta = 0.5$, corresponding to Fig.~\ref{fig:local}(e).
We also list the top 10 countries that have the highest values
in out-degree, in-degree, out-strength, and in-strength.
}
\begin{tabular}
{c | l | l | l | l | l | l }
\hline\hline
Rank & first impact & second impact  & out-degree & in-degree & out-strength & in-strength \\
\hline
1   &    United States (US)               &   US                 &   US                       &       UK            &     US                &     US  \\
2   &    France (FR)           &   UK                 &   UK                       &       US            &     FR            &     UK  \\
3   &    United Kingdom (UK)               &   Hong Kong (HK)                 &   FR                   &       RU        &     UK            &     HK  \\
4   &    Germany (DE)          &   Luxembourg (LU)          &   DE                  &       CH   &     DE           &     LU  \\
5   &    Italy(IT)            &   FR             &   NL          &       FR        &     IT             &     FR   \\
6   &    Japan (JP)            &   DE            &   IT                    &       Panama island (PA)        &     JP             &     DE  \\
7   &    The Netherlands (NL) &   CH        &   JP                    &       DE       &     CN             &     RU   \\
8   &    China (CN)            &   Russia (RU)             &   CA                   &       LU    &     NL   &     CH  \\
9   &    Switzerland (CH)      &   Brazil (BR)             &   CH              &       Australia (AT)    &     CH       &     BR  \\
10  &    Canada (CA)           &   Singapore (SG)          &   Libya  &       SG     &     AT           &     SG   \\
\hline\hline
\end{tabular}
\label{tab:iks}
\end{table*}

\section{Local Spreading Model}
\label{sec:local}

\subsection{Model Description: Local Spreading}
\label{subsec:localmodel}

In the present study, we propose two different models for the spread of shocks. In the first model for the local
spreading, we focus on the initial spread of crisis by considering the shock transmission procedure upto two
steps:
\begin{itemize}
\item As the original source of the banking crisis, the country $i$ is picked sequentially one by one.
\item Step I (the first wave): Due to the country $i$'s domestic banking crisis, $i$ is forced to reduce its banks
abroad by the fraction $\alpha$. The bigger is $\alpha$, the more serious is the original
crisis.
\item  Step II (the second wave): Each country ($j$) where $i$ has its banks (i.e., $w_{ij} > 0$) realizes this
change of number of banks within country, and thus tries to pull back its banks in $i$ to reduce
the future risk. We assume that the
country $j$ tries to sell the assets it has in $i$ very quickly, and thus there is an
inevitable reduction of total values. We take into account this reduction by introducing
another parameter $\beta$ so that $j$ closes down all the banks in $i$ and then moves the fraction
$1-\beta$ of banks to other country $k$.  The larger is $\beta$, the more loss the country $j$ has in $i$.
For the choice of $k$, we choose the country in which
$j$ has the most banks (except for $i$). If there are more than two countries with the maximum number of banks,
we choose the one at random.
\end{itemize}

For the directed and weighted networks, one can define the out-strength $s^{\rm out}$ and
the in-strength $s^{\rm in}$
of each vertex $i$ as follows:
\begin{eqnarray}
s^{\rm out}_i & \equiv & \sum_{j=1}^N w_{ij}, \\
s^{\rm in}_i & \equiv & \sum_{j=1}^N w_{ji},
\end{eqnarray}
which simply correspond to the total number of overseas banks $i$ built,
and the total number of foreign banks in $i$, respectively.
We also note that the total number of  foreign banks in the world satisfies
$B = \sum_i s^{\rm out}_i = \sum_i s^{\rm in}_i$.

Within the local spreading model in this Section, the size of global crisis caused by the country $i$ is
measured by the change of numbers of banks in the whole world, after each step
(I and II). We denote $B^{(0)}$ as the total number of foreign banks before any crisis,
and $B_i^{(1)}$ [$B_i^{(2)}$] is the total number of foreign banks after the first
(the second) wave in Step I (II) caused by the original crisis in $i$. The seriousness
of the first and the second waves is measured by the quantity we call ``impact'' $I$
defined by the change of number of banks in the world:
\begin{eqnarray}
I^{(1)}_i \equiv B^{(0)} - B_i^{(1)},  \\
I^{(2)}_i \equiv B_i^{(1)} - B_i^{(2)} .
\end{eqnarray}

\subsection{Impact Sizes: Local Spreading}
\label{subsec:localimpacts}

Figure~\ref{fig:local} displays the first impact and the second impact caused by
each country. It is clearly seen that the sizes of impacts are very different
from country to country and that as the first and the second waves become
more serious ($\alpha$ and $\beta$ become larger) the impact sizes also increase.
We also list the ten countries that have the highest values in $I^{(1)}$ and $I^{(2)}$
in Table~\ref{tab:iks} for $\alpha = \beta = 0.5$. The top countries which
cause the biggest first impacts are somehow different from those countries
with the biggest second impacts. Specifically, HK, Luxemberg, and
Singapore are well known financial centers although they do not have the
biggest economy like US. The list in Table~\ref{tab:iks} implies that
the countries with the big first impact and the countries with the big second
impact can have very different characteristics. In order to elaborate this,
we compare local node properties (in- and out-degrees and strengths) with
the impact sizes in Fig.~\ref{fig:local}.
We observe clear positive correlation  between the strengths and the impacts
and   conclude that the
first impact is mostly determined by the out-strength, while the second impact
by the in-strength. As a matter of fact, the observed strong correlations between
the strengths and the impacts are not surprising at all, in view of our local
spreading model of banking crisis in Sec.~\ref{subsec:localmodel}: The first wave (Step I)
only deals with the overseas bank of $i$ and thus the reduction of the number
of banks are mostly determined by the $w_{ij}$. On the other hand, the second
wave (Step II) is heavily influenced by how many foreign banks exist in $i$
and thus it should be closely related with $w_{ji}$.
In detail, the dependence of $I_i^{(1)}$ on
$w_{ij}$ can be understood as follows: The number of banks $B_i^{(1)}$ after
the first wave is written as
$B_i^{(1)} = \sum_{jk} w_{jk}^\prime$, where $w_{jk}^\prime$ is the value after Step I.
Since the crisis occurred at $i$, we get
\begin{eqnarray}
B_i^{(1)} &= &\sum_{j\neq i, k}w_{jk} + \sum_{k}w_{ik}^\prime,  \nonumber \\
&= &B^{(0)} - \sum_k w_{ik} + \sum_k \lfloor(1- \alpha)w_{ik}\rfloor,
\end{eqnarray}
where $\lfloor x \rfloor$ is the floor of $x$, i.e., the largest integer
not greater than $x$. We then obtain
\begin{equation}
I_i^{(1)} = s_i^{({\rm out})} - \sum_k\lfloor(1- \alpha)w_{ik}\rfloor.
\end{equation}
We note that as $\alpha \rightarrow 1$, $I_i^{(1)} \rightarrow s_i^{({\rm out})}$
as also confirmed in our simulations. When $\alpha$ is small but nonzero, one
can also understand $I_i^{(1)} \approx k_i^{({\rm out})}$ since
$\lfloor(1- \alpha)w_{ik}\rfloor  = w_{ik} -1$.  The similar reasoning can also
be applied for Step II to understand $I_i^{(2)} \approx s_i^{({\rm in})}$
for large $\beta$ and  $I_i^{(2)} \approx k_i^{({\rm in})}$ for small $\beta$.
For intermediate values of $\alpha$ and $\beta$, the detailed numbers of banks
must be taken into account to understand what happens in the bank network for
Step I and II.

In Table~\ref{tab:iks}, we also list
the 10 countries which rank the highest in terms of out-degree, in-degree,
out-strength, and in-strength. Although the top 10 list for the first impact
and the out-strength (and for the second impact and the in-strength) look
almost identical, there still exist small differences.

\subsection{Emergence of a New Superpower: Local Spreading}

One can define the superpower in our network in
different ways. In terms of the out-degree, the out- and
the in-strengths, US is the superpower as can be seen
in Table~\ref{tab:iks}. On the other hand, in terms of
the in-degree, UK has the highest power. London has been
playing the financial center of the world for a long time,
but our data indicate that UK is the superpower only in terms
of the number of countries that had branches in UK. In terms
of the number of banks that are located within a country,
US is the most important country. We next study what will happen
if one country becomes the source of crisis and thus collapses.
From the simulation of our model, it is found that if the origin
of the crisis is US, UK becomes the financial center in terms
of the in-degree and the in-strength. On the other hand,
if the crisis first occurs in France, not US, but UK becomes
the superpower in terms of in-strength. In all other cases,
US keeps its position as superpower.

\section{Global Spreading Model}
\label{sec:global}
\subsection{Model Description: Global Spreading}
\label{subsec:globalmodel}
Our local spreading model in Sec.~\ref{sec:local} is not really able to capture
the iterative nature of shock transmission across the global bank network.
In the present Section, we introduce a global spreading model of shocks,
in which the banking crisis originating from the source country
spreads to other countries as follows:

\begin{itemize}
\item  A country $i$ is picked sequentially as the origin of the banking crisis.
The country $i$ undergoes banking crisis and thus reduces the number of
banks by the fraction $\alpha$ in neighbor countries (initiation of crisis).
This step is identical to our local spreading model in Sec.~\ref{subsec:localmodel}.

\item The country $j$ from which $i$ reduced banks in above step calculates
the loss of incoming banks between the initial and the current states.
If the fraction is larger than $\gamma$ ($\in [0,1]$)
the country $j$ realizes that crisis is coming and also withdraws its banks abroad
by the fraction $\alpha$ in the same way as the country $i$ did in the above
initiation step. $\alpha$ and $\gamma$ are the two parameters in the present
model of global spreading of shocks; the former controls how severe the crisis is
and the latter works as a threshold for the detection of the crisis. For simplicity
the same values of $\alpha$ and $\gamma$ are assigned for all countries, respectively.

\item The above step of reduction of banks is performed iteratively and thus banking
crisis propagates across the global banking network. The process stops when
no more withdrawals of banks occur. This iterative nature makes the present global
spreading model different from the local spreading model in Sec.~\ref{sec:local}.
If $\gamma$ is set to unity, the crisis stops propagating beyond $i$'s
neighbor countries, and the outcome becomes identical to the Step I in Sec.~\ref{sec:local}.
As $\gamma$ is decreased from unity, countries become more fragile against baking crisis.
The impact of the source country $i$ on the global outcome of banking crisis is
measured similarly to Sec.~\ref{sec:local}, and we denote it as
$I^{(g)}_i$ with the superscript $g$ meaning global spreading:
\begin{equation}
\label{eq:Ig}
I^{(g)}_i \equiv B^{(0)}-B^{(g)}_i,
\end{equation}
where $B^{(0)}$ is the total number of banks before crisis
and $B^{(g)}_i$ is the number of surviving banks in the world after the global transmission
of the shock originating from $i$ stops.
\end{itemize}

\subsection{Impact Sizes: Global Spreading}

\begin{figure}[t]
\includegraphics[width=0.45\textwidth]{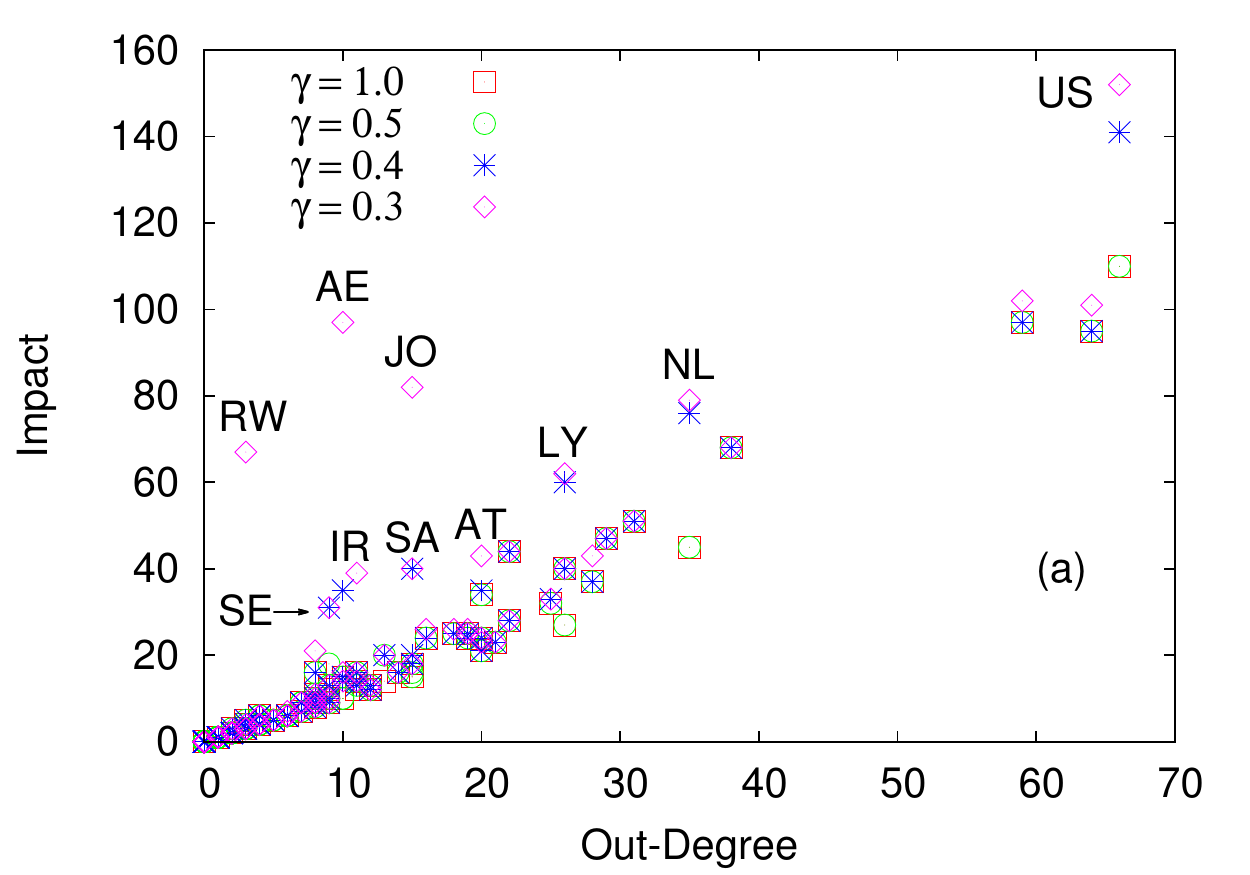}
\includegraphics[width=0.45\textwidth]{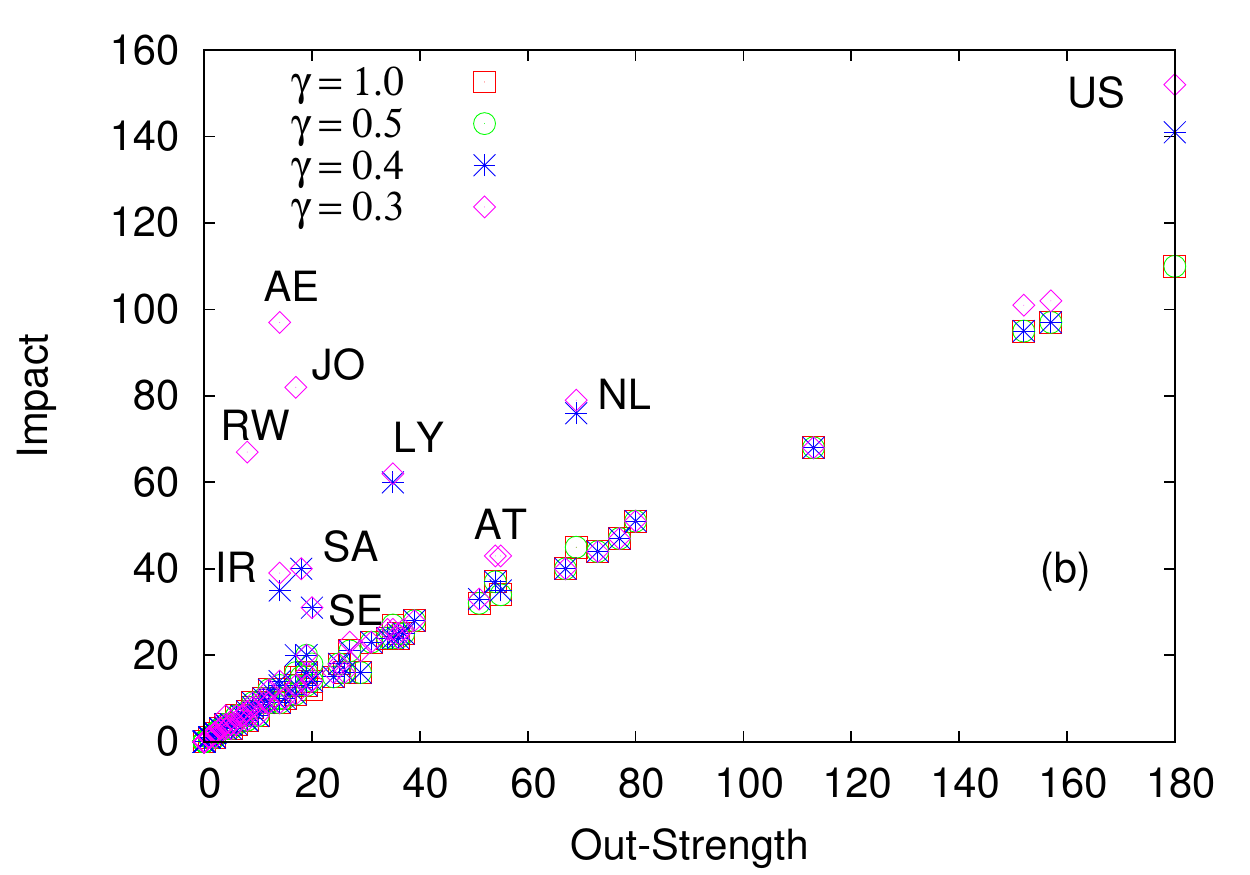}
\caption{(Color online)
Global spreading model: Scatter plot for the impact ($I^{(g)}_i$) versus (a)
out-degree and (b) out-strength, both for  $\alpha=0.5$. When $\gamma$ is
larger than $0.5$, impact is approximately linear with respect to (a) the
out-degree and (b) the out-strength.  For small values of $\gamma$,
some countries exhibit strong deviations from the linear relations.
Outliers that have impact values deviating much from linearity are marked
(see Table~\ref{tab:global} for the symbols of those countries).}
\label{fig:global}
\end{figure}

We simulate the spread of crisis varying $\gamma$ from $1.0$ to $0.3$ for $\alpha = 0.5$.
As $\gamma$ becomes smaller than 0.3, all countries become fragile unanimously
without showing significant differences. We also note that for $\gamma \gtrsim 0.5$,
the impact values do not strongly depend on $\gamma$ any longer.
Figure~\ref{fig:global} shows the impact $I^{(g)}_i$ in Eq.~(\ref{eq:Ig})
caused by the crisis originating from the country $i$ with
respect to (a) the out-degree and (b) the out-strength of $i$
at $\alpha=0.5$. We observe that impacts are strongly correlated with
the out-degrees and the out-strengths. Although the general increasing
tendency of impact seen in Fig.~\ref{fig:global} is an expected behavior (see
discussion in Sec.~\ref{subsec:localimpacts}),
the deviation from the simple linear relation observed for some countries
for $\gamma < 0.5$ demands a better understanding.
In order to elucidate the origin of these outliers, we compute the
difference in impact values between $\gamma = 0.3$ and $\gamma = 1.0$:
\begin{equation}
\label{eq:deltaI}
\Delta I_i \equiv I^{(g)}_i(\gamma = 0.3) - I^{(g)}_i(\gamma = 1.0),
\end{equation}
which measures  additional network effect on top of the
first step of direct crisis spreading only to neighbor countries.
[Note that $I^{(g)}_i(\gamma = 1.0)$ has the same value as $I^{(1)}_i$ in Sec.~\ref{subsec:localmodel}.]
Consequently, the outliers in Fig.~\ref{fig:global} are those countries with
large values of $\Delta I_i$. The top 10 countries with the highest $\Delta I_i$
are listed in Table~\ref{tab:global}, where we also include
the total number $N_{\rm crisis}$ of endangered countries in which the loss
fraction has exceeded $\gamma$ during the crisis propagation.
When the crisis is initiated from a country with $\Delta I_i \approx 0$, the crisis
does not propagate beyond the country's direct neighbors and become localized.
We believe that our measure $\Delta I_i$ can properly capture how big a systemic
risk the country $i$ can cause in the global bank network.

When the crisis starts from $i$, $i$ reduces the banks from its direct
neighbors ($j$'s) by the fraction $\alpha$, and thus the number of banks
$j$ loses is written as  ${\lceil \alpha w_{ij} \rceil}$, where $\lceil x \rceil$
is the ceiling function returning the smallest integer not less than $x$.
When the loss fraction ${\lceil \alpha w_{ij} \rceil}/{\sum_{k} w_{kj}}$
is larger than $\gamma$, the country $j$ also reduces its banks from other
countries. Since $\lceil x \rceil < x + 1$, the above condition that $j$
also reduces its banks from other countries is written as
\begin{equation}
\frac{\gamma \sum_{k\neq i} w_{kj} - 1}{\alpha - \gamma} < w_{ij},
\label{eq:cond1}
\end{equation}
for $\alpha > \gamma$.
From Eq.~(\ref{eq:cond1}), $j$ is expected to be vulnerable to the crisis originating from $i$,
if $j$ has incoming banks only from a few neighbors. It is to be emphasized that
the condition~(\ref{eq:cond1}) is only a proxy since it does not reflect the crisis
propagating from countries other than from the origin. In actual computer simulation of
our global spreading model, crisis can spread beyond direct neighbors of the crisis
origin.

\begin{table}[t]
\caption{Global spreading model: Top $10$ countries of the largest
impact difference $\triangle I$ [see Eq.~(\ref{eq:deltaI})
and Fig.~\ref{fig:global}] are listed.
The number $N_{\rm crisis}$ of countries which are brought into crisis
from the shock originating from those 10 countries are also listed.
}
\begin{tabular}{c|c|c}
\hline\hline
Name & $\Delta I $ & $N_{\rm crisis}$ \\
\hline
United Arab Emirates (AE) & 87 & 17 \\
Jordan (JO) & 66 & 18 \\
Rwanda (RW) & 62 & 18  \\
United States (US) & 42 & 12  \\
Libya (LY) & 35 & 15  \\
The Netherlands (NL) & 34 & 8  \\
Iran (IR) & 26 & 6  \\
Saudi (SA) & 25 & 2  \\
Sweden (SE) & 13 & 6  \\
Austria (AT) & 9 & 4  \\
\hline\hline
\end{tabular}
\label{tab:global}
\end{table}

\begin{table}[t]
\caption{Global spreading model: The most vulnerable 10 countries
with influence within top 20\%. The values of
vulnerability ($\max_k w_{ki}/\sum_j w_{ji}$)
and influence ($\sum_j w_{ij}$) together with the number $N_{\rm loops}$
of feed-forward loops which has the country as the root
are listed (see text for more details).}
\begin{tabular}{c|c|c|c}
\hline\hline
Name & vulnerability & influence & $ N_{\rm loops}$ \\
\hline
Togo & 1.0 & 27 & 10 \\
Hong Kong & 0.54 & 19 & 28 \\
Kuwait & 0.5 & 36 & 85 \\
Greece & 0.5 & 26 & 27 \\
Sweden & 0.5 & 20 & 11 \\
South Africa & 0.5 & 19 & 8 \\
Colombia & 0.43 & 15 & 14 \\
Japan & 0.375 & 77 & 223 \\
Libya & 0.34 & 35 & 64 \\
India & 0.34 & 31 & 57 \\
Portugal & 0.3 & 20 & 14 \\
\hline\hline
\end{tabular}
\label{tab:vulnerable}
\end{table}

What is the origin of the outlying behavior
(the big impact difference $\Delta I$ in Fig.~\ref{fig:global})?
In order to answer this,  we introduce the concept of ``the trigger country'' as
follows: When the crisis from a source country $i$ reaches a trigger country
$j$, the country $j$ plays the role of the trigger that although $j$ does not
directly contribute in the impact difference $\Delta I_i$ of the source
country $i$, $j$ does so indirectly by relaying the crisis to other countries,
eventually bringing  many other countries into crisis. Our definition of
the trigger country leads to the two conditions the trigger country must
satisfy: First, it is vulnerable to the crisis of its incoming neighbors
(vulnerability).
Second, it can affect other countries so that its outgoing neighbors can
also be in crisis (influence). The vulnerability of the country $i$
can effectively be measured by $\max_k w_{ki}/\sum_j w_{ji}$, which
is proportional to the risk that $i$ is driven to crisis by one of its
incoming neighbors $k$. The influence, the second ingredient to be a trigger,
can be measured by $i$'s outgoing strength $\sum_j w_{ij}$. We expect
that these two local properties to gauge vulnerability and influence
can capture the possibility that a country can be a trigger, although
it is based on the assumption that most contribution to $\Delta I_i$
comes from $i$'s nearest and next-nearest neighbors.
In Table~\ref{tab:vulnerable}, the most vulnerable $10$ countries
of which their values of influence rank within top 20\% are listed.
See Saudi in Table~\ref{tab:global} for example. Saudi ranks high
by provoking the trigger country Kuwait (Kuwait is the only one
out-going neighbor of Saudi that has become endangered). Kuwait in
Table~\ref{tab:vulnerable} has higher values in both vulnerability and
influence, which leads us to conclude that our two measures, vulnerability and
influence, can effectively capture the possibility for a country to become a
trigger country. For the spread of crisis originating from United Arab Emirates,
Jordan, and Rwanda, it is found that Libya and Togo play the role of the trigger
countries, whereas Sweden plays a role of the trigger for US and the Netherlands
(see Table~\ref{tab:vulnerable}). 

To summarize, the outliers in
Fig.~\ref{fig:global} and Table~\ref{tab:global} are either countries
which have triggers as their directly connected neighbors (like United
Arab Emirates, Jordan, and Rwanda all with the trigger country Togo as
their neighbors),  or trigger countries themselves (like Libya).
We also emphasize that a trigger country may not be an outlier in 
Fig.~\ref{fig:global} and Table~\ref{tab:global}: The influence in
Table~\ref{tab:vulnerable} is defined as the outgoing strength, and
consequently takes into account only the first wave impact. In
contrast, outliers are defined in such a way that only the impact beyond the
first wave is measured [see Eq.~(\ref{eq:deltaI})]. Outliers are 
those countries which exhibit big network effect beyond the direct
connections, and we believe triggers are conceptually useful to
understand why some countries become outliers.

We also observe that even when a country $k$ does not satisfy the condition~(\ref{eq:cond1}) 
and thus survives the first step of shock transmission
from the origin country, it might fall into crisis in the next step by the crisis
propagating from other country.
Suppose that a country $j$ which has banks of
the country $i$ ($w_{ij} > 0$) opens branch banks in another neighbor country
$k$ ($w_{jk} > 0$), in which $i$ also has branch banks ($w_{ik} > 0$).
In this situation, it is possible that the crisis from the country $i$ affects $k$ both
through the direct path and the indirect path via the country $j$, which is a
typical example of the so-called ``feed-forward loop'' with $i$ as the root.
If  crisis from a source country and a trigger country arrive at the target country
through the feed-forward loop, the target country is more likely to fall into crisis.
We measure the number $N_{\rm loops}$ of the feed-forward loops and also include them in
Table~\ref{tab:vulnerable}.

\begin{figure}[t]
\includegraphics[width=0.49\textwidth]{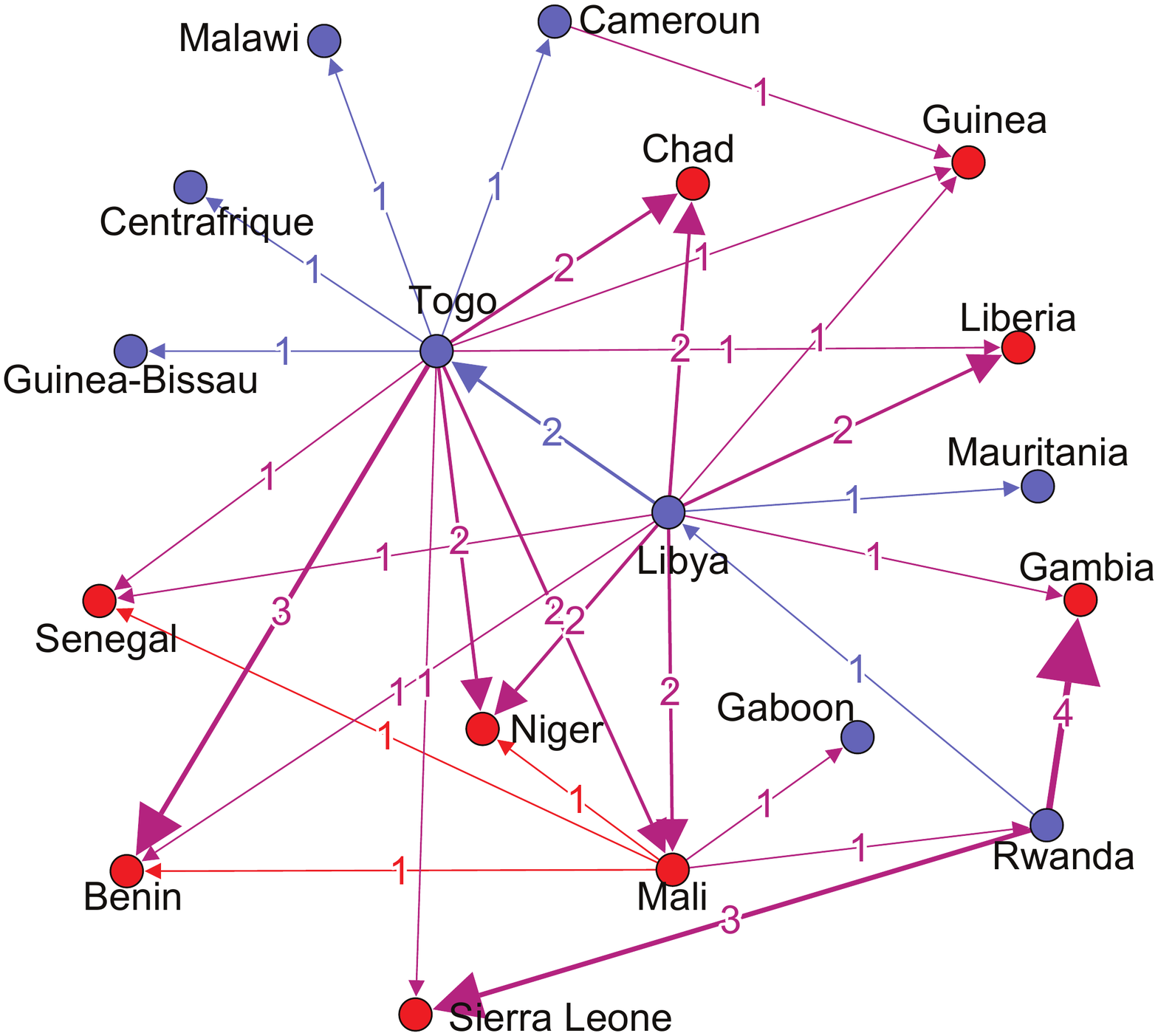}
\caption{(Color online) Global spreading model: The subnetwork of crisis spread
with Rwanda as the crisis origin ($\alpha=0.5$ and $\gamma=0.3$). Each country
undergoes crisis through either the direct mechanism [the
condition~(\ref{eq:cond1}) satisfied] or the feed-forward-loop mechanism (see text for
details).  The numbers on the directed edges are the weights (the number of
overseas banks between two countries).  The feed-forward-loop mechanism
eventually affects nine countries, Gambia, Liberia, Guinea, Chad, Senegal,
Niger, Mali, Benin, and Sierra Leone, all colored red. The blue-colored countries
are the ones for which the direct crisis condition~(\ref{eq:cond1}) is satisfied.
In this subnetwork of crisis, Libya and Togo are the trigger countries (see text).
}
\label{fig:Rwanda}
\end{figure}

To illustrate the process of spreading crisis through feed-forward loops, we
plot the subnetwork of spreading crisis which starts from Rwanda in
Fig.~\ref{fig:Rwanda} for $\alpha=0.5$ and $\gamma=0.3$. Among the nearest
neighbors of Rwanda, the only country which satisfies Eq.~(\ref{eq:cond1})
to become endangered is Libya.
Not only Libya contributes much for the increase of the
impact value of Rwanda due to its large value of influence (see  Table~\ref{tab:vulnerable}),
it also plays an important role by composing feed-forward loops:
Rwanda and Libya compose feed-forward loops together with Gambia and Mali.
Although the crisis from Rwanda affects Gambia and Mali in the first step,
these two survive because the condition~(\ref{eq:cond1}) is not fulfilled.
However, in the second step of shock transmission when Libya affects these two,
they eventually collapse into crisis from the feed-forward loop mechanism.
Eventually, all nine countries, Gambia, Liberia, Guinea, Chad, Senegal, Niger, Mali, Benin,
and Sierra Leone become affected by the crisis originating from Rwanda through the
the feed-forward loops with Libya and Togo playing important roles (see Fig.~\ref{fig:Rwanda}).
We find that all eight countries except Saudi and Austria in Table~\ref{tab:global}
become the victim of the crisis spread by the feed-forward-loop mechanism.

Feed-forward loops can significantly increase the impact of crisis beyond
the simple sum of the impact of local crisis from the trigger and the
the source countries.
Without any feed-forward loops, it is easily seen that
Libya must have $\Delta I = 21$, instead of the actual value $\Delta I = 35$.
From this, we conclude that the big impact difference $\Delta I$ in global
spreading comes from combination of two factors: large out-strength and
feed-forward loops. Especially, the effect of feed-forward loop from the
trigger country increases the number of endangered countries significantly.
Consequently, we conclude that the existence of trigger countries
with feed-forward loops needs to be taken into account for
forecasting the pattern of crisis spreading across the whole bank network.
We note that the number of feed-forward loops is not the only indicator
for the big scale of crisis spread. For example, although Japan has
233 feed-forward loops,  many of its neighbors are not vulnerable
and thus they work as cushion not as trigger for the crisis spread.

\section{Policy Implications}
First, from the data, this research shows that for most of countries in the
world, their banking service accounts are not balanced. Some have larger
in-strength than out-strength, such as UK. These countries have bank service
trade deficit because they import more than export. On the contrary, other
countries have bank service surplus as they have larger out-strength than
in-strength. Whether a country has bank service surplus or deficit is a result
of their trade policy implemented. If countries decide to implement an
export-promote-trade policy, such as Japan in 70s and 80s, they will end up
with bank service trade surplus, otherwise it is going to be deficit.
This research shows that the sequential shocks have different impacts on these
two types of countries. Export-oriented countries need to prepare more
``capital cushion'' as they are going to be exposed to the initial shock;
however the import-oriented countries also need to prepare enough capital for
the second round of the shock.

Second, as lender of last resorts, governments are expected to step in and
provide safety nets for the crisis attacked banking system. The size of the
financial aid however is, in general, not easy to determine. This research
indicates that the variation of the shock sizes determine the different impacts
on the banking system measured by the reduction number of the banks. The
magnitude of the reduction can provide certain information on the seriousness
of the effect, hence certain implication for the size of the bailout packages.

Third, our results for the global spreading model suggest that in
order to prevent systemic risk from spreading identification of countries
with a big network effect is important. We also believe that our
concept of the trigger countries can be important for international
intervention to stop crisis spreading: The trigger countries can trigger 
a big scale of risk spreading, but if they are provided some
international financial support in early stage of risk propagation, 
they can also work as a bottle neck for spreading, enhancing the
stability of the whole international baking network.

\section*{Acknowledgment}
X. F. was supported by Leading Academic Discipline Project of Shanghai
Municipal Education Commission J51206; National Education Commission Project
Number 11YJA790030. Shanghai Social Science Fund Project Number 2011BJB015,
Shanghai Municipal Education Commission Innovation Project，Project
Number 12ZS166, and  National Post Doctor Fund 2011M5000786.
B. J. K. acknowledges the support from the National Research
Foundation of Korea (NRF) grant funded by the Korea government (MEST) (Grant
No. 2011-0015731), and the support from the Asia Pacific Center for
Theoretical Physics.
Feedbacks from the 5th Shanghai
International Symposium on Nonlinear Sciences and Applications 2012 are
appreciated. X. F.  would like to thank Xiaoyan Huang, Jingjie Gu for
research assistance.

\bibliographystyle{apsrev4-1}
\bibliography{bank}

\end{document}